# A controlled ac Stark echo for quantum memories

B. S. Ham

Center for Photon Information Processing, and School of Electrical Engineering and Computer Science,
Gwangju Institute of Science and Technology
123 Chumdangwagi-ro, Buk-gu, Gwangju 61005, S. Korea

bham@gist.ac.kr

**Abstract**
A quantum memory protocol of controlled ac Stark echoes (CASE) based on a double rephasing photon echo scheme via controlled Rabi flopping is proposed. A double rephasing scheme of photon echoes inherently satisfies the no-population inversion requirement for quantum memories, but the resultant absorptive echo has been a fundamental problem. In the present studies, firstly, the first echo in the double rephasing scheme is dynamically controlled not to affect the second echo by using unbalanced ac Stark shifts. Secondly, a control Rabi pulse coherently manipulates the second echo to be emissive via controlled coherence conversion. Owing to far-off resonance of ac Stark fields, the present protocol can be applied for wavelength-selective dynamic control of quantum memories in real time for erasing, buffering, and multiplexing.

**Introduction**

To overcome population inversion constraint in photon echoes for quantum memory applications, several modified photon echo schemes have been developed since 2001. Of them are controlled reversible inhomogeneous broadening [1-5], atomic frequency comb (AFC) [6-8], silent echoes [9,10], dc Stark echoes [11,12], controlled double rephasing (CDR) [13,14], and optically locked photon echoes [15]. Although the population inversion constraint has been lifted off in these protocols, the intrinsic absorptive echo in a double rephasing scheme [9-11] has been a fundamental drawback [14]. The solution for the absorptive echo in a double rephasing scheme has already been discussed for controlled coherence conversion (CCC) via optical Rabi flopping to a third state [13-15]. Here, we present a new quantum memory protocol of controlled ac Stark echoes (CASE) based on double rephasing photon echoes, where unbalanced ac Stark fields work for silent echo formation for the first echo not to affect the second echo [9-12], and a controlled Rabi flopping works for coherence inversion for the second echo to be emissive. The most important benefit of using ac Stark fields is in dynamic control of quantum information processing in real time. Compared with dc Stark echoes [11,12] technological convenience such as in cavity coupling difficulties or impractical conditions for phase mismatching [9] are additional advantages.

Recently demonstrated ac Stark modulations in conventional two-pulse photon echoes [16] have shown the same mechanism as in the dc Stark echoes for the atom phase control [11,12], in which the Stark field gives an on-purpose phase shift (turbulence) to the transient atoms not to rephase for the first echo. Thus, combining double ac Stark modulations with controlled Rabi flopping in the double rephasing photon echo scheme induces an inversion-free, emissive second echo via silencing the first echo: CASE. With an advantage of bigger ac Stark splitting, CASE offers technological advances compared to dc Stark echoes [11,12] or phase mismatching [9,10] in rare-earth doped solids. Because rare-earth doped-solids have a low electric susceptibility [14] requiring high electric fields, optical ac Stark fields are naturally compatible with a short-pulse photon echo scheme. Thus, the present CASE can be a good candidate for nano-photonics-based quantum devices as preliminary demonstrated in nano-cavity photon echoes [17].

The characteristics of CASE are as follows. First, a gradient field required in dc Stark echoes is just optional due to a large phase shift $\Phi_{AC}$, and spectral preparation is not required either. Second, with a large optical detuning $\Delta_{AC}$, the ac Stark pulse duration $\tau$ can be short enough: $\Phi_{AC} \sim \Delta_{AC}\tau$. This means that the benefit of ultrafast all-optical information processing of photon echoes is inherited. Third, the intensity of the ac Stark field can be as week as possible owing to the bigger generalized Rabi frequency $\Omega'$ ($\sim \Delta_{AC}$). Fourth, the detuning selectivity of the ac Stark fields renders multiple channel accessibility in frequency domain, as in the wavelength division multiplexing of fiber-optic communications, where this feature can offer channel manipulations in multimode quantum memories. Fifth, an unknown quantum state to be stored or to be retrieved can be dynamically manipulated for an eraser [16] or entangled photon generations [18]. Sixth, CASE is insensitive to spectral tailoring such as in free induction decay and short-storage time. Seventh, the storage time can be extended as long as the spin population decay time [15].

To solve the absorptive echo drawback in a double rephasing scheme [9-11], the CDR echo protocol has recently been intensively analyzed for a near perfect quantum memory [14]. As the general concepts of double rephasing photon echoes are shown in the Supplementary Information (see Fig. S1), there is no way to obtain an emissive photon echo in a double rephasing scheme without population inversion, yet. The echo observations in refs 9-11, however, are due to nonuniform pulse area applied to each atom resulting from a Gaussian pulse shape in a transverse spatial mode perpendicular (x and y axis) to the beam propagation direction (z axis) as well as Beer's law-dependent absorption along the longitudinal axis. The echo observations in refs. 9-11 are not artifacts, but lower echo efficiency is inevitable due to photon losses. Unlike the original $\pi-\pi$ control Rabi flopping in a Doppler medium for emissive photon echoes [1], solids media behave differently resulting in an absorptive echo due to coherence inversion [2,14]: $\rho(t) \to -\rho(t')$. In that sense, the controlled AFC echo [19] belongs to the absorptive echo category, too [20].

**Results and Discussion**

Before the discussion of CASE, we analyze *firstly* on-demand atom phase control by ac Stark fields to erase the first echo, and *secondly* CCC by a control Rabi pulse to make the absorptive second echo emissive. Here the control Rabi pulse whose pulse area is $2(2n-1)\pi$ is resonant between the excited state and an auxiliary ground state. For numerical simulations, we solve nine time-dependent density matrix equations obtained by Liouville-von Neumann equations under rotating wave approximations for an inhomogeneously broadened optical three-level system interacting with multiple resonant optical pulses [21]. The CASE is an extended version of the ac Stark modulation in a single rephasing scheme [16] to the double rephasing scheme combined with CCC. The phase shift by an ac Stark field (AC) is given by $\Phi_{AC} = \Omega'\tau$, where $\Omega' = \sqrt{\Omega_{AC}^2 + \Delta_{AC}^2}$, and $\Omega_{AC}$ and $\tau$ are the Rabi frequency and pulse duration of AC. Then it may be asked *what $\Phi_{AC}$ is required for the silent echo e1 not to affect e2*. Here it should be noted that individual atom phase evolutions decay with respect to optical phase decay time, whereas macroscopic (or overall) coherence decays with respect to the inverse of inhomogeneous width of the system.

To answer this question, we start with dressed states induced by the ac Stark field. In a two-level system composed of a ground state $|1\rangle$ ($\omega_1$) and an excited state $|2\rangle$ ($\omega_2$), the interaction Hamiltonian with an ac Stark field results in dressed states as shown in Fig. 1(a): $|2\pm\rangle$ ($\omega_\pm$) $= \omega_b + \frac{1}{2}\Delta'_{AC} \pm \frac{1}{2}\Omega'$. Here, this optical system must be inhomogeneously broadened by $\Delta_{inh}$ to satisfy the photon echo condition. Thus under the ac Stark field the individual atom detuning is denoted by $\Delta'_{AC} = \Delta_{AC} + \delta_j$, where $\delta_j$ is the detuning of the $j^{th}$ atom in $\Delta_{inh}$ from the line center. Because $\Delta_{inh}$ is symmetric, we treat the system as a collection of symmetrically detuned atom pairs $\pm j^{th}$ across the line center in addition to a resonant atom (zeroth) located at line center. The generalized Rabi frequency of AC is expressed by $\Omega' = \Delta'_{AC}\sqrt{1+\left(\frac{\Omega_{AC}}{\Delta'_{AC}}\right)^2}$. For $\Delta_{AC} \gg \Omega_{AC}; \Delta_{inh}$, it is approximately $\Omega' \simeq \Delta'_{AC}$, $\omega_+ \simeq \omega_b + \Delta'_{AC}$, and $\omega_- \simeq \omega_b$. Thus, the Rabi oscillation of ac Stark field is dominated by the AC detuning, $\Delta_{AC}$.

### A. ac Stark modulation: echo erasing

Figure 1(b) is the pulse sequence of the ac Stark modulations, and Figs. 1(c)~1(f) are the corresponding results. Because the second echo e2 is a macroscopically rephased coherence of the first echo e1, e2 must be affected unless e1 is not erased. To modify the doubly rephased photon echo in Fig. 1(b), there are two important tasks to do with the atom phase control. *First, echo e1 must be silent as to not affect echo e2. Second, echo e2 must be emissive to be radiated out of the medium.* To begin with we analytically derive how D-excited atoms' evolutions are described by using time-dependent density matrix equations under rotating wave approximations. The key evolution parameter in coherent transients like photon echoes is the detuning $\delta_j$ in $\Delta_{inh}$: $\rho_{12}(t) \propto \rho_{12}(0)e^{\pm i\delta_j}$. By the inserted AC in Fig. 1(b), a phase shift $\Phi_{AC}(\tau)$ is added to each D-excited atom coherence: $e^{\pm i\delta_j} \to e^{\pm i\delta_j + i\Phi_{AC}(\tau)}$. Here the sign of $\Phi_{AC}(\tau)$ is predetermined by AC detuning, where '+' ('-') represents a blue (red) detuning. This means that the inhomogeneously broadened atoms are simply frequency shifted by $\Delta'_{AC}$, and the phase evolution of each atom is accelerated by $\Omega'$ during $\tau$.

Figures 1(c) and 1(d) show the ac Stark-induced photon echo erasing for $\Phi_{AC} = \Omega'\tau = \pi/2$. Figures 1(e) and 1(f) are for individual atoms corresponding to Figs. 1(c) and 1(d), respectively. The phase addition of $\pi/2$ by AC to the system coherence $\rho_{12}(t)$ results in swapping between real and imaginary parts of $\rho_{12}(t)$ (To compare with a bare double rephasing scheme, see Supplementary Information Fig. S1): $e^{\pm i\delta_j} \to e^{\pm i(\delta_j + \pi/2)}$. Thus, all $Im\rho_{12}$ (absorption) becomes zero at e1 timing as shown in Figs. 1(c) and 1(e), while $Re\rho_{12}$ (dispersion) becomes maxima in coherence as shown in Figs. 1(d) and 1(f). Here individual atom phase evolutions are not affected by the macroscopic coherence e1 as shown in Fig. 1(e). The first task of silent echo e1 is achieved now with $\Phi_{AC} = \pi/2$. In other words, the unbalanced ac Stark field applied to a single rephasing scheme can erase echoes dynamically in real time.

To understand the photon echo erasing mechanism by AC in Fig. 1(c), detailed analyses are performed in Fig. 2. In Fig. 2(a), the AC Rabi frequency $\Omega_{AC}$ varies for a fixed pulse duration $\tau$ from the top to the bottom (see e1): $\Phi_{AC} = \frac{\pi}{5}; \frac{\pi}{4}; \frac{\pi}{3}; \frac{\pi}{2}(center); \frac{2\pi}{3}; \frac{3\pi}{4}; \pi$. For this, the atom's inhomogeneous broadening is given in Fig. 2(b) as applied to all figures. To analyze the results in Fig. 2(a), a simple model is used for the e1 amplitude efficiency $\eta$ as a function of $\Phi_{AC}$, and the results are shown in Figure 2(c):

$$\eta(e1) \propto \cos\Phi_{AC}, \quad (1)$$

where $\eta$ (solid curve) oscillates as a function of AC pulse duration $\tau$ for a fixed $\Omega'$: $I_{e1} \propto \frac{1}{2}(1 + \cos 2\Phi_{AC})$ [22]. The dashed curve is for the resonant atoms ($\delta_j=0$; $\Delta'_{AC} = \Delta_{AC}$) as a reference. The gradual reduction of $\eta$ is due to random phases among coherently excited atoms due to detuning $\delta_j$, as observed in free induction decay. However, the echo e1 erased for $\Phi_{AC} = (2n-1)\pi/2$ does not matter with the pulse area. This means that a big $\Phi_{AC}$ offers less sensitivity in echo erasing (silent echo) condition. Figure 2(d) summaries the resultant phase evolution speed of each atom: The more the atom is blue-detuned, the more its phase evolution accelerates. As $\Phi_{AC}$ increases, the gap of phase evolution speed widens, resulting in enhanced decay in $\eta$ as shown in Fig. 2(c).

### B. ac Stark echoes

Figures 3(a) and 3(b), respectively, show two different cases of unbalanced (asymmetric) and balanced (symmetric) ac Stark fields applied to CASE. If AC2 turns on before (after) e1 and $\Omega_{AC2}=\Omega_{AC1}$, then it is called 'balanced' ('unbalanced'). Figures 3(c) and 3(d) are the corresponding results. Figures 3(e) and 3(f) are the details of Figs. 3(c) and 3(d) respectively for individual atoms. As shown in Figs. 3(c) and 3(e), the unbalanced CASE results in erasing (silence) of e1, but an absorptive echo e2. For the analysis of this silent echo e1, the atom phase evolution for a symmetrically detuned pair at $\pm\delta_j$ is viewed as followings for the pulse sequence of D → AC1 → R1: $e^{\pm i\delta_j t} \to e^{\pm i\delta_j t + i\Phi_{AC1}} \to e^{\mp i\delta_j T - i\Phi_{AC1} \pm i\delta_j t'}$; $t' = t - T$; $T = t_{R1} - t_D$; $t_j$ stands for the arrival time of pulse $j$. Thus, all $Im\rho_{12}$ become zero at $t' = T$ (t = 2T) for echo e1 by $\Phi_{AC1} = \frac{\pi}{2}$ as shown in Fig. 3(e). The phase addition by AC1, however can be completely compensated by AC2 if $\Phi_{AC2} = \Phi_{AC1}$. The atom phase evolution for R1 → AC2 → R2 is $e^{\pm i\delta_j(t'-T) - i\Phi_{AC1}} \to e^{\pm i\delta_j(t'-T) - i\Phi_{AC1} + i\Phi_{AC2}} \to e^{\mp i\delta_j(T+T'-T) \pm i\delta_j t''} = e^{\pm i\delta_j(t''-T')}$, where $T' = t_{R2} - t_{e1}$; $t'' = t - (T' + 2T)$. Thus, at $t'' = T'$ (or $t = 2(T'+T)$), echo e2 is generated. The echo e2 is of course absorptive, which is useless for applications. On the other hand, turning AC2 on before e1 in Fig. 3(b) makes no change as shown in Fig. 3(d) (see also Supplementary Information Fig. S1(b)). This is exactly due to a complete cancellation of the phase shift by AC2 via R1 rephasing before e1 formation: $e^{\pm i\delta_j t + i\Phi_{AC1}} \to e^{\mp i\delta_j T - i\Phi_{AC1} + i\Phi_{AC2} \pm i\delta_j t'} = e^{\mp i\delta_j(t'-T)}$; $t' = t - T$.

The absorptive echo e2 in Fig. 3(c) can also be analyzed another way. As discussed in Fig. 1, the π/2 pulse area of the ac Stark pulse swaps atom coherence between the real and imaginary components in the density matrix elements. In the double rephasing photon echoes with unbalanced ac Stark fields as in Fig. 3(a), the atom coherence control by each optical pulse in Fig. 3(a) is described in the pulse sequence of D → AC1 → R1 → AC2 → R2:

$$i\rho(t_D) \to r\rho(t_{AC1}) \to [r\rho(t_{R1})]^* \to [i\rho(t_{AC2})]^* \to i\rho(t_{R2}), \quad (2)$$
$$r\rho(t_D) \to i\rho(t_{AC1}) \to [i\rho(t_{R1})]^* \to [r\rho(t_{AC2})]^* \to r\rho(t_{R2}), \quad (3)$$

where $t_j$ stands for the time right after the pulse $j$. Here $r$ ($i$) stands for real (imaginary). Obviously echo e2 has the same form as the D-excited coherence, resulting in an absorptive echo.

### C. CASE: Controlled ac Stark echoes

To convert the absorptive echo e2 into an emissive echo, the present quantum memory protocol CASE is introduced in Fig. 4. As shown in Fig. 4(a), a control 2π pulse C is added right after the second rephasing pulse R2, where the function of the control pulse is an optical Rabi flopping between the excited state |2⟩ and an auxiliary state |3⟩ forming a three-level system. The mechanism of the control pulse has been intensively studied for coherent transients [1,2,13-15,19], where the control pulse plays coherence inversion (see Fig. 6 in ref. 14): $\rho_{12}(t_{R2}) \to -\rho_{12}(t_C)$. Thus, the absorptive echo in Eqs. (1) & (2) turns out to be emissive by the 2π C as shown in Figs. 4(b) and 4(d): $i\rho(t_{R2}) \to -i\rho(t_C)$. The echo e2 has exactly the same form as the two-pulse photon echoes: $i\rho(t_D) \rightarrow [i\rho(t_R)]^* = -i\rho(t_R)$. The Bloch vector model of coherent transients for the control $2\pi$ pulse has also been discussed numerically in refs. 13 and 14. Figure 4(c) shows all individual atom phase evolutions for Fig. 4(b), where e1 is completely erased by AC1 ($\Phi_{AC1} = \pi/2$) due to $Im\rho_{12} \leftrightarrow Re\rho_{12}$ as discussed in Fig. 3(c), while echo e2 gets a maximum coherence for emission. Figure 4(d) shows the details (extension) of Fig. 4(c) for a detuned atom at $\delta_j$=150 kHz. At t=17.1 μs the control $2\pi$ pulse C turns on for 0.2 μs (see the shaded area). By C, the coherence of both $Re\rho_{12}$ and $Im\rho_{12}$ is completely reversed. For the echo e2, system population is of course the same as that for the data absorption due to the double rephasing. The large detuning of ac Stark fields negligibly affects population redistributions (Supplementary Figure S2). Here, the control $2\pi$ pulse can be divided into two $\pi$ pulses with an advantage of storage time extension as well as backward echo propagation, where coherence conversion between optical and spin transitions is involved [23]. For the backward echo scheme, the retrieval efficiency can be near unity [1,23]. Thus, the absorptive echo problem in the modified photon echoes using a double rephasing scheme [9~11] as well as in the controlled AFC in a single rephasing scheme [19] can be fixed by adding the control $2\pi$ (or $\pi-\pi$) pulse at the end.

**Methods**
**Numerical calculations.** For the numerical calculations presented in all figures, all decay rates are set for zero for the visualization of individual atom-phase evolutions, otherwise specified. The density matrix approach is powerful in dealing with an ensemble medium for light-matter interactions especially to trace coherence evolutions of individual atoms, so that phase evolutions can be visualized clearly for coherent transients such as photon echoes. For the analytic approach, a simple phase term of each atom under rotating wave approximation appeared in the Hamiltonian is used to show how each atom's phase evolves with the interacting optical fields in time domain. The equations of motion of the density matrix operator ρ are determined by Liouville-von Neumann equations[21]:

$$\frac{d\rho}{dt} = \frac{i}{\hbar}[H, \rho] - (decay\ terms), \quad (4)$$

where Hamiltonian *H* is under rotating wave approximation:

$$H = \begin{bmatrix} \delta_1 & \Omega_1 & 0 \\ \Omega_1 & \delta_2 & \Omega_2 \\ 0 & \Omega_2 & 0 \end{bmatrix}, \quad (5)$$

where $\Omega_1$ ($\Omega_2$) is the Rabi frequency of the resonant optical pulse between states $|1\rangle$ and $|2\rangle$ ($|3\rangle$ and $|2\rangle$). Each optical pulse is assumed a square pulse for simple calculations. The state $|3\rangle$ is an auxiliary ground state in Fig. 1(a). The $\delta_1$ ($\delta_2$) is a detuning of each atom group from the resonant optical pulse $\Omega_1$ ($\Omega_2$), where it is given by an optical inhomogeneous broadening. The pulse D, R1, R2, AC1, and AC2 are for $\Omega_1$, and the ac Stark detuning $\Delta_{AC}$ is added to $\delta_1$ (see Section B). The pulse C in Fig. 4 is for $\Omega_2$. Then equations (4) and (5) results in:

$$\dot{\rho}_{ij} = -\frac{i}{\hbar}\sum_k(H_{ik}\rho_{kj} - \rho_{ik}H_{kj}) - \frac{1}{2}(\gamma_{ik}\rho_{kj} + \rho_{ik}\gamma_{kj}), \quad (6)$$

where γ is the decay term. For simplicity γ is set zero. For all calculations, inhomogeneously broadened optical system (850 kHz in FWHM) is chosen, where the optical inhomogeneous broadening and the interacting optical Rabi frequencies are chosen similar to the experimental parameters in a rare-earth doped $Pr^{3+}$ doped $Y_2SiO_5$ (Pr:YSO). The actual optical and spin decay rates of Pr:YSO are negligibly small similar to kHz or less compared to the optical inhomogeneous broadening. Here the actual 4 GHz optical inhomogeneous broadening in Pr:YSO can be manipulated experimentally to be narrowed to be ~ 1 MHz or less easily via spectral hole-burning owing to the three hyperfine states in the ground level. The inhomogeneous broadening is assumed Gaussian in the calculations. Nine total time-dependent density matrix equations obtained from equation (6) are completely solved numerically without approximations. In each figure, the 99.55% of atoms in the optical inhomogeneous broadening is divided into 201 groups ($\delta_j$) at 10 kHz spectral spacing, then each atom group is calculated in time domain with a different weight factor and a different detuning $\delta_j$ determined by the Gaussian profile, and finally all atom groups are summed up together for overall coherence. Each optical pulse duration is set 0.1 μs. The time increment for the calculations in equation (6) is 0.01 μs.

**Conclusion**
In summary a new quantum memory protocol of controlled ac Stark echoes (CASE) is presented in a double rephasing photon echo scheme with an unbalanced ac Stark pulse pair and a control Rabi pulse, where the control Rabi pulse converts the absorptive echo into emissive one via coherence inversion. Using time-dependent density matrix equations we numerically solved CASE without approximations and also analytically discussed phase evolutions of the photon echo system. The control $2\pi$ Rabi pulse can also be applied for storage time extension as well as near unity echo efficiency by dividing it into a π−π control pulse pair. The present CASE quantum memory protocol can be applied to various applications of wavelength division multiplexing for dynamic control of real time quantum information processing.

**Acknowledgment**

This work was supported by ICT R&D program of MSIP/IITP (1711028311: Reliable crypto-system standards and core technology development for secure quantum key distribution network) and GIST-Caltech Program in 2016.

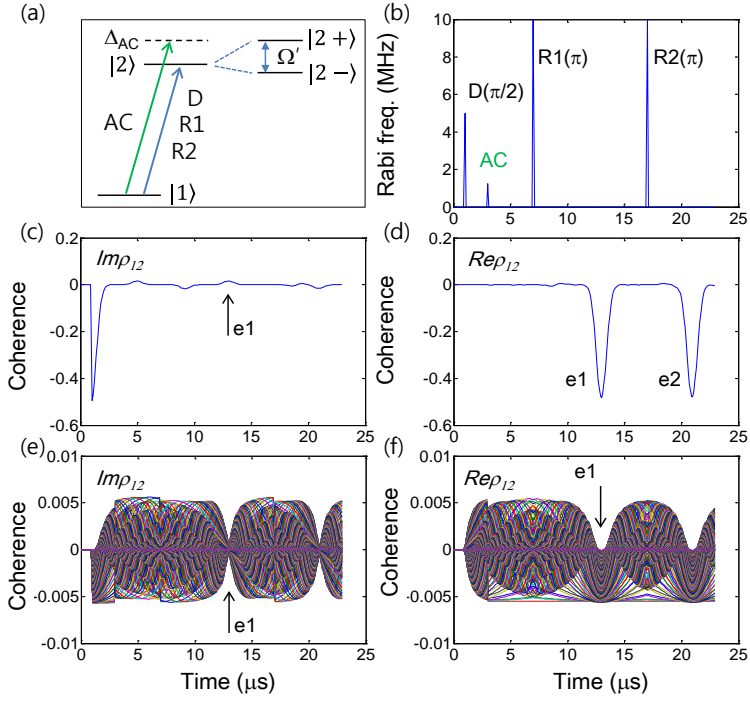

**Figure 1. Echo erasing by ac Stark turbulence.** (a) Energy level diagram for AC Stark modulation: ac Stark (AC), Data (D), Rephasing (R1), and Rephasing (R2). (b) Pulse sequence for (a). (c)-(f) Numerical calculations for optical coherence based on time-dependent density matrix equations. For this the optical inhomogeneous width is divided into 201 groups of atoms at 10 kHz space. (c) and (d) are for the sum of all atoms, where (e) and (f) are for all individual atoms. Optical inhomogeneous broadening of the two-level system is 850 kHz (FWHM). All decay rates are set zero. Initially all atoms are in the ground state: $\rho_{11}(0)=1$. Each pulse duration is 0.1 μs. The time of arrival of D, AC, R1 and R2 are 1, 3, 7, and 17 μs, respectively. AC Rabi frequency is $\Omega_{AC}=1.25$ MHz, and its detuning is $\Delta_{AC}=\sqrt{15}\Omega_{AC}$. The generalized Rabi frequency of the AC Stark field is $\Omega' = \sqrt{\Delta_{AC}^2 + \Omega_{AC}^2}$. The e1 and e2 are echoes by R1 and R2, respectively.

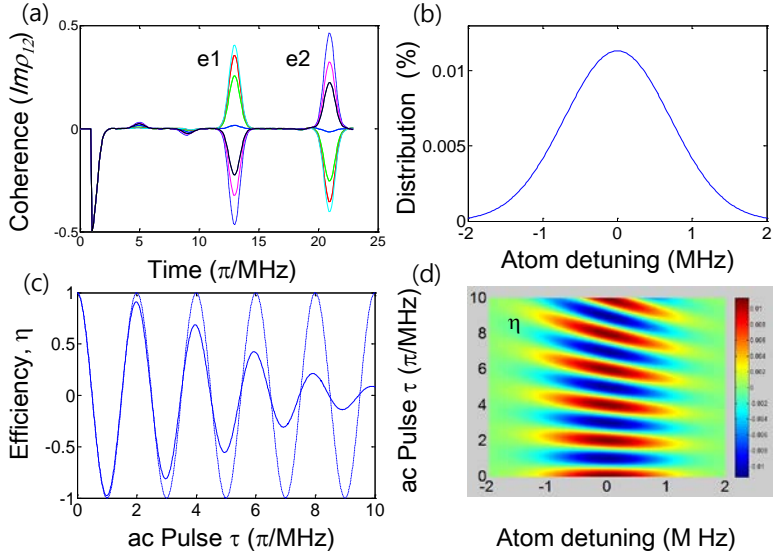

**Figure 2. ac Stark influence to photon echoes.** (a) ac Stark photon echoes for different ac Stark Rabi frequency, $\frac{\Omega_{AC}}{N}$ with a fixed pulse duration of 0.1 μs (for schematics and parameters, see Fig. 1). Cyan (N=5; π/5); Red (N=4; π/4); Green (N=3; π/3); Blue (N=2; π/2); Black (N=3/2; 2π/3); Magenta (N=4/3; 3π/4); Dashed (N=1; π). All decay rates are zero. $\rho_{11}$(t=0)=1. Blue curve (center) is a reference for a π/2 ac Stark pulse area. The e1 and e2 are echoes by R1 and R2, respectively. (b) Optical inhomogeneous broadening of 850 kHz (FWHM) in (a) has a Gaussian distribution. (c) Photon echo (e1) amplitude efficiency η vs. ac Stark pulse duration τ. Generalized ac Stark Rabi frequency is Ω′=1.0 MHz. Dotted (Solid): For homogeneous (inhomogeneous) atoms. (d) A 2D color map of (c) for Gaussian distributed atoms. See the text for the details of calculations.

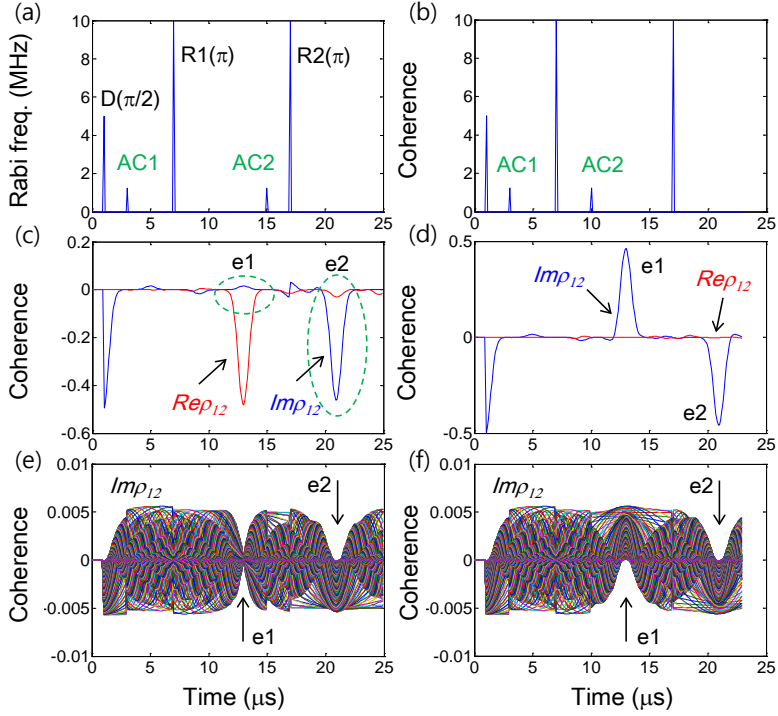

**Figure 3. AC Stark control in doubly rephased photon echoes.** (a), (b) Pulse sequence: Data (D), Rephasing (R1), Rephasing (R2), AC Stark pulses (AC1) and (AC2). (c)-(f) Numerical calculations for optical coherence based on time-dependent density matrix equations. (c) and (e) are for (a), where (d) and (f) are for (b). The time of arrival of D, AC1, R1, and R2 are 1, 3, 7 and 17 µs, respectively. The arrival time of AC2 is (a) 15 µs and (b) 10 µs. Other parameters are the same as in Fig. 1, otherwise specified.

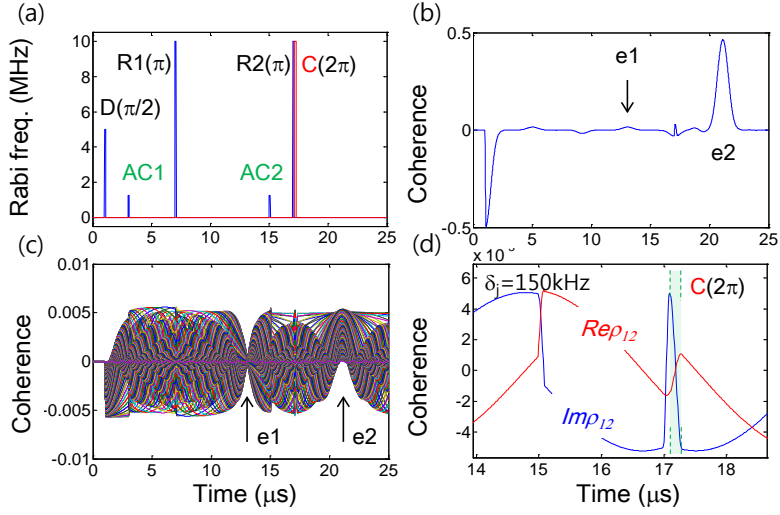

**Figure 4. Controlled ac Stark echoes (CASE) in doubly rephased photon echoes.** (a) Pulse sequence: Data D, Rephasing R1, Rephasing R2, ac Stark pulses AC1 & AC2, and control pulse C. (c)-(d) Numerical calculations for optical coherence based on time-dependent density matrix equations. The arrival time of C is 17.1 μs with 0.2 μs pulse duration. All other parameters are same as those in Fig. 3. All parameters are the same as in Fig. 3, otherwise specified.

# Supplementary Information

Figure S1 shows numerical simulations of double rephasing photon echoes. Figure S1(a) is the pulse sequence, and Figs. S1(b)~S1(f) are the corresponding results. The double rephasing photon echo scheme inherently provides two echoes e1 and e2, and the final echo e2 is under no population inversion. To get echo e2, however, echo e1 must be silent, otherwise it affects echo e2. With a backward propagation scheme for both R1 and R2, a silent echo e1 was demonstrated [9], where the second propagation direction of the echo e2 is forward. Although echo e2 is under no population inversion, the echo e2 is absorptive as shown in Fig. S1(b). Thus, the echo e2 cannot be radiated out of the optically dense medium. Making e2 emissive is the main task of the controlled coherence conversion [20]. It has been discussed in the controlled double rephpasing photon echoes [13].

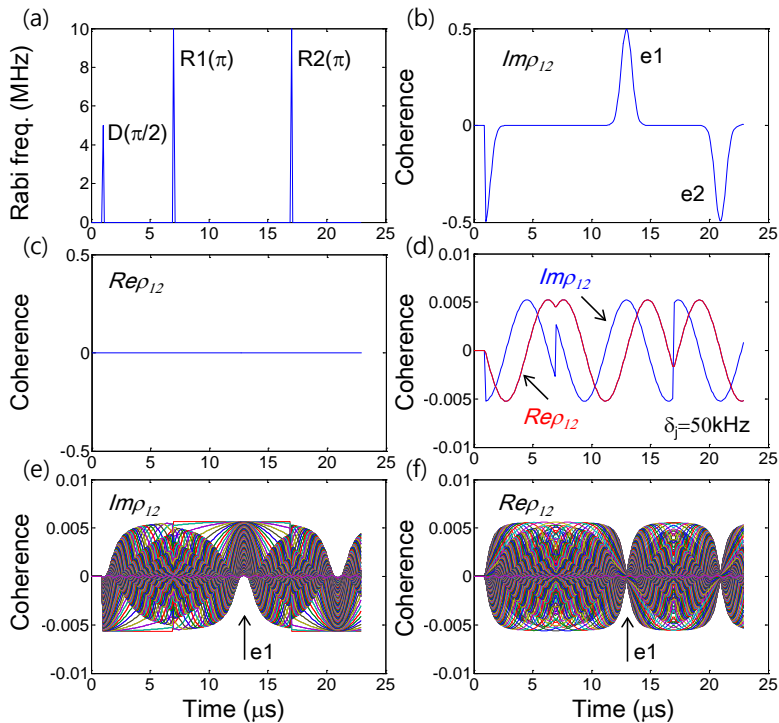

**Figure S1. Doubly rephased photon echoes.** (a) Pulse sequence: Data D, Rephasing R1, and Rephasing R2. (b)-(f) Numerical calculations for optical coherence based on time-dependent density matrix equations. (b) and (c) are for sum of all atoms, where (e) and (f) are corresponding all individual atoms. All other parameters are the same as in Fig. 1.

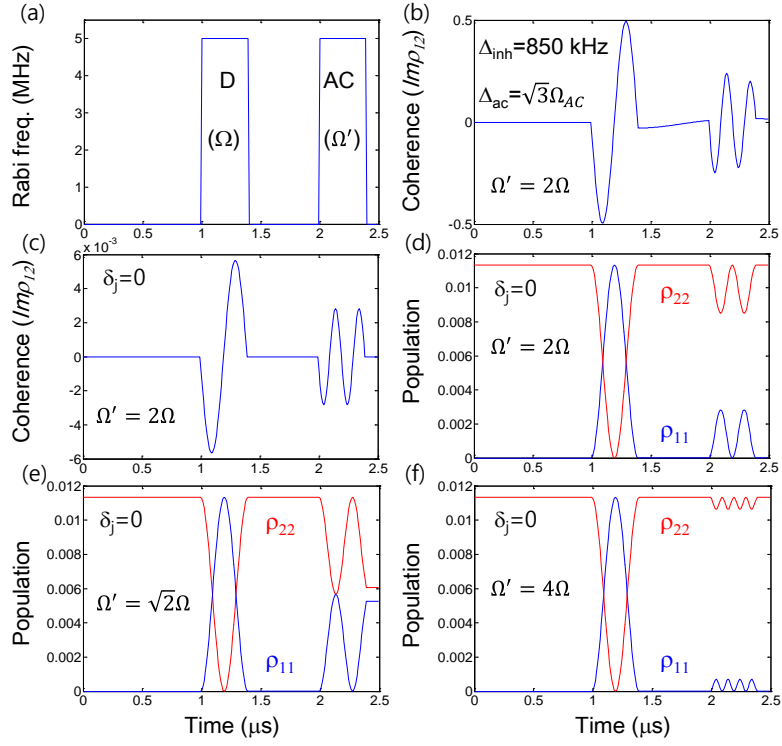

**Figure S2. Rabi oscillations in a two-level system.** (a) Pulse sequence of resonant (D) and detuned (AC) excitations. Optical inhomogeneous broadening of the two-level system is 850 kHz (FWHM). A detuning $\Delta_{AC}$ for the off-detuned case is $\Delta_{ac}=\sqrt{3}\Omega_{AC}$. The Rabi frequency of D is 5 MHz corresponding to a $2\pi$ pulse area. All decay rates are set zero for simplicity. Initially all atoms are on the ground state $|1\rangle$: $\rho_{11}=1$. D and AC are independent. (b) Rabi oscillation of each system. (c) Rabi oscillation for only resonant atom at line center. (d) Population oscillations on the ground state (blue) and excited state (red) for (c). (e) and (f) Population oscillations for (e) $\Delta_{AC}=\Omega_{AC}$ and (f) $\Delta_{AC}=\sqrt{15}\Omega_{AC}$. Generalized Rabi frequency for an ac field AC is given by $\Omega' = \sqrt{\Delta_{AC}^2 + \Omega_{AC}^2}$. All data are obtained from numerical calculations of time-dependent density matrix equations.